\documentclass[twocolumn,showpacs,showkeys]{revtex4}
\pdfoutput=1
\usepackage{graphicx}

\usepackage{bm}

\begin{document}

\title{Numerical simulations of possible finite time singularities in
  the incompressible Euler equations: comparison of numerical methods}

\author{Tobias Grafke}
\affiliation{Institute for Theoretical Physics I, 
             Ruhr-Universit\"at Bochum, Germany}
\author{Holger Homann}
\affiliation{Institute for Theoretical Physics I,
             Ruhr-Universit\"at Bochum, Germany}
\author{J\"urgen Dreher}
\affiliation{Institute for Theoretical Physics I, 
             Ruhr-Universit\"at Bochum, Germany}
\author{Rainer Grauer}
\email{grauer@tp1.rub.de}
\affiliation{Institute for Theoretical Physics I,
             Ruhr-Universit\"at Bochum, Germany}

\begin{abstract}
  The numerical simulation of the 3D incompressible Euler equation is
  analyzed with respect to different integration methods. The
  numerical schemes we considered include spectral methods with
  different strategies for dealiasing and two variants of finite
  difference methods. Based on this comparison, a Kida-Pelz like
  initial condition is integrated using adaptive mesh refinement and
  estimates on the necessary numerical resolution are given. This
  estimate is based on analyzing the scaling behavior similar to the
  procedure in critical phenomena and present simulations are put into
  perspective.
\end{abstract}



\maketitle

\section{\label{intro}Introduction}
The question, whether the incompressible Euler equations develop
singularities in finite time starting from smooth initial conditions,
remains an outstanding open problem in applied mathematics. Although
substantial progress has been made in recent years using a more
geometrical viewpoint
\cite{constantin-fefferman-etal:1996,gibbon:2002, gibbon:2007,
  deng-hou-etal:2005, deng-hou-etal:2006}, it is yet not clear from
numerical simulations, whether the assumptions of the theorems for
non-blow up are fulfilled for flows evolving from simple smooth
initial conditions. Singular structures, evolving in finite time or
simply ``fast enough'', may play a similar role as shock-like
structures in compressible flows, providing structures which dominate
the energy dissipation even in the non-viscous situation (see Eyink
\cite{eyink:1994, eyink-sreenivasan:2006, eyink:2007} and references
therein).

In this paper, we study a Kida-Pelz like flow with different numerical
schemes: spectral methods with different strategies of dealiasing
(this extends the study of Hou and Li \cite{hou-li:2007} and confirms
their results), two finite difference methods and a finite volume
method.  Studying the structures of vorticity, it turns out that the
differences between the various methods of dealiasing are more
pronounced than between the spectral methods and the finite
difference/volume methods. This result suggests that resolving the
vorticity structures is more important than the order of the numerical
scheme. It also justifies the use of finite difference/volume methods
in adaptive mesh refinement (AMR) simulations to resolve the vorticity
structures.

Using AMR simulations up to an effective resolution of $4096^3$ mesh
points and comparing the results to lower resolution runs, we observe
that the standard way of presenting a $1/|{\bm \omega}|$ plot in time
may lead to misleading conclusions. However, looking at normalized
plots reveals the issue of numerical resolution in a convincing manner.

\section{\label{numerics}Numerical schemes}

In this section we compare spectral methods with different dealiasing
and finite difference/volume methods.

\subsection{Spectral methods and dealiasing}
We use a standard spectral method where the time stepping is performed
with a strongly stable 3rd-order Runge-Kutta method
\cite{shu-osher:1988} in Fourier space and where nonlinearities are
calculated in real-space. On Linux-clusters, the FFTW-library is used
whereas the library P3DFFT \cite{pekurovski-yeung-etal:2006} from the
San Diego Supercomputer Center is used on the IBM Regatta series and
on BlueGene/L.

We use three ways of dealiasing the spectral data:
\begin{enumerate}
\item Spherical mode truncation: this is used in turbulence
  simulations 
  (Biskamp and M\"uller \cite{biskamp-mueller:1999}). The spherical
  mode truncation puts a sphere of radius $\frac{N}{2}$ in Fourier
  space and nullifies all modes outside this sphere.
\item Standard 2/3 rule: same as above, but using a radius of
  $\frac{2}{3}\frac{N}{2} = \frac{N}{3}$
  \cite{canuto-hussaini-etal:1987}. This is the most common way of
  dealiasing spectral data.
\item High-order exponential cut-off: this method was introduced by
  Hou and Li \cite{hou-li:2007} and consists of introducing a
  high-order exponential filter function $\rho(k) = \exp(-\alpha
  (|k|/N)^m)$ with $\alpha = 36$ and $m=36$.
\end{enumerate}

\subsection{Finite difference/volumes methods}
All presented finite difference/volume methods are second order and
use the same strongly stable 3rd-order Runge-Kutta method
\cite{shu-osher:1988} as used in the spectral simulations.

We implemented three different versions of real-space methods:
\begin{enumerate}
\item Staggered grid formulation of Harlow and Welsh
  \cite{harlow-welsh:1965}: Normal components of the velocity are
  located at their respective cell faces and the pressure is defined
  at the cell centers. This allows a exact Hodge-decomposition such
  that no pressure oscillations occur. In addition, it conserves
  momentum and energy and could thus also been seen as a finite volume
  method.
\item Vorticity formulation for AMR: From our previous AMR studies
  \cite{grauer-marliani-etal:1998,grauer-marliani:2000} we know that
  the coarse-fine grid interpolations are very sensitive in the 3D
  Euler simulations. As in the former simulations we choose to perform
  all data exchange and interpolation using the vorticity ${\bm
    \omega} = \nabla \times \mathbf{u}$. Here, the vorticity is
  defined at cell centers and we applied a tri-cubic interpolation for
  coarse-fine grid interpolation. Then, three Poisson equations are
  solved for the cell-centered vector Potential $\mathbf{A}$ and
  staggered values for the velocity $\mathbf{u} = \nabla \times
  \mathbf{A}$ are obtained.
\item Finite volume method: this method is similar to the former but a
  finite volume method \cite{bell-colella-etal:1989,
    bell-marcus:1992b, grauer-marliani-etal:1998} is used instead of
  finite differences.
\end{enumerate}

\subsection{Comparison}

\begin{figure}[!b]
  \includegraphics[width=1\columnwidth]{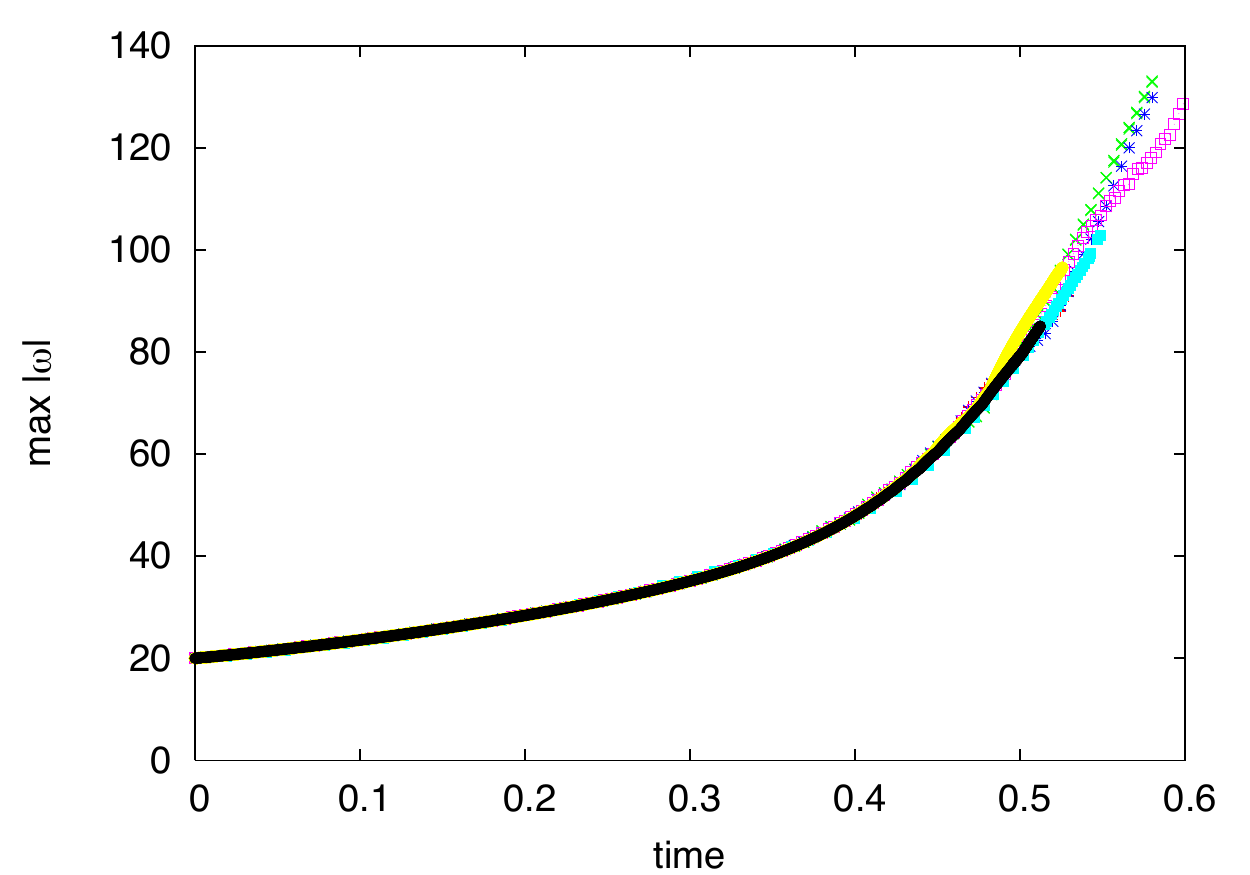}
  \caption{\label{growth_omega} Growth of $\max |{\bm \omega}|$ 
    for all implemented numerical schemes.}
\end{figure}

\begin{figure}[ht!]
  \includegraphics[height=0.179\textheight]{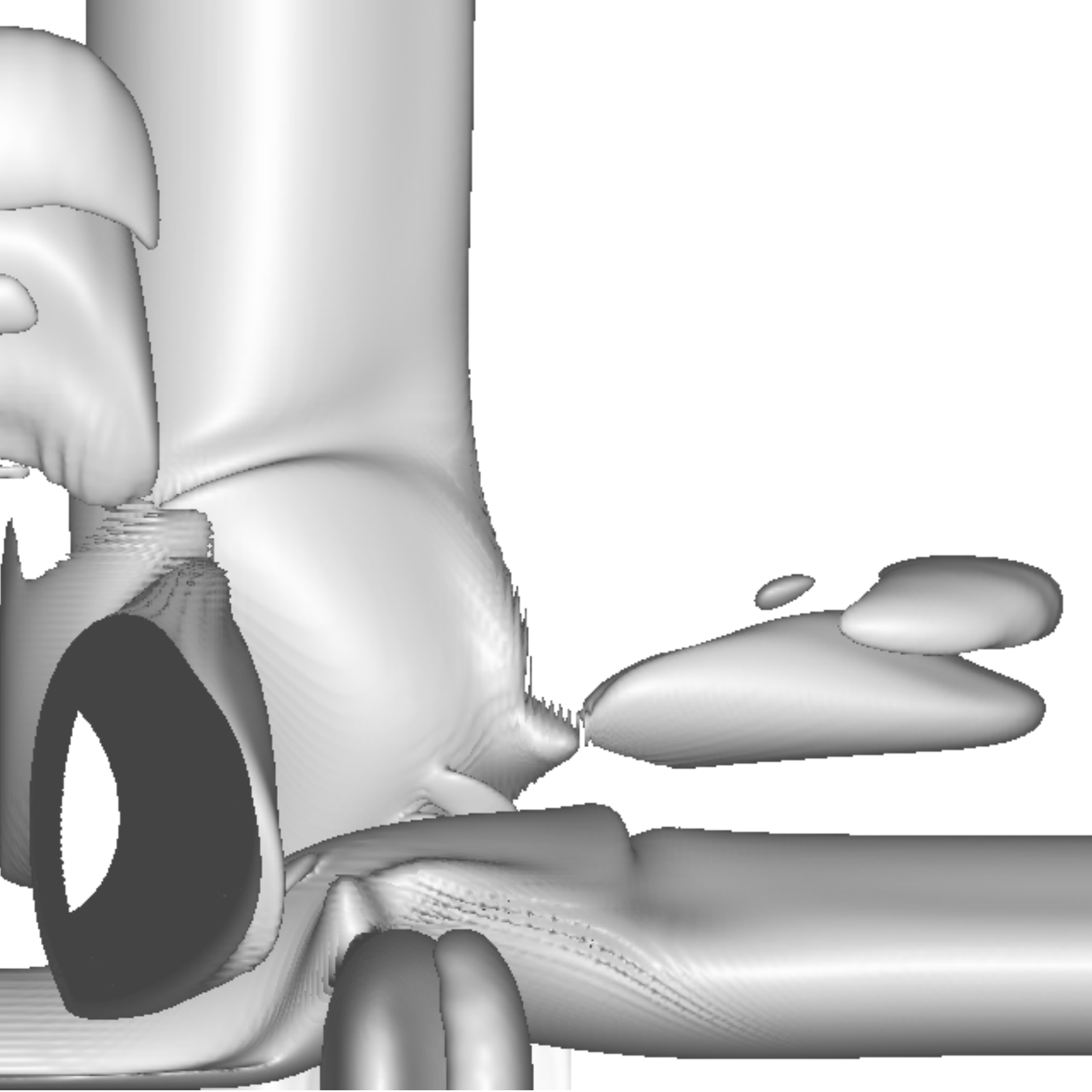} \\
  \vspace*{2mm}
  \includegraphics[height=0.179\textheight]{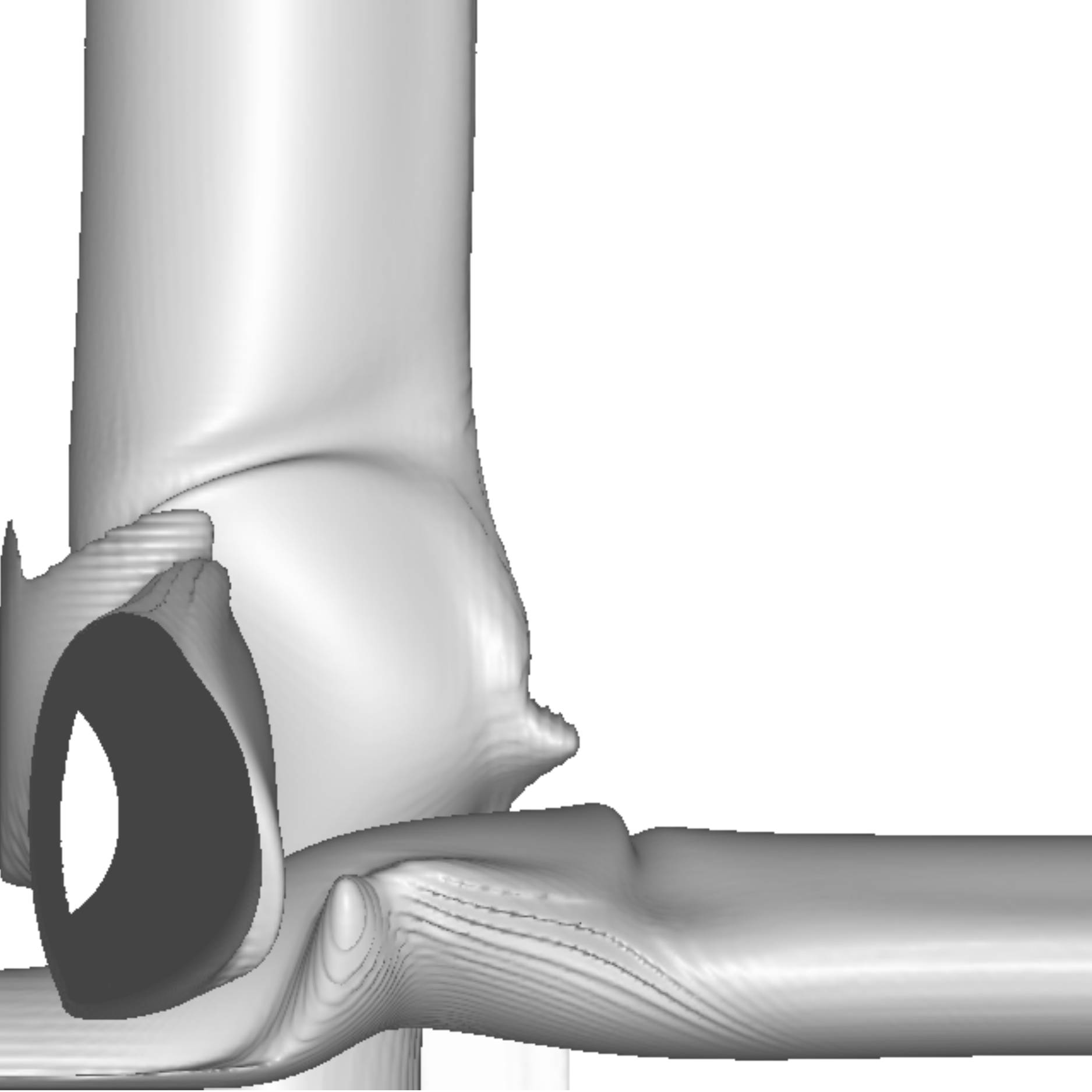} \\
  \vspace*{2mm}
  \includegraphics[height=0.179\textheight]{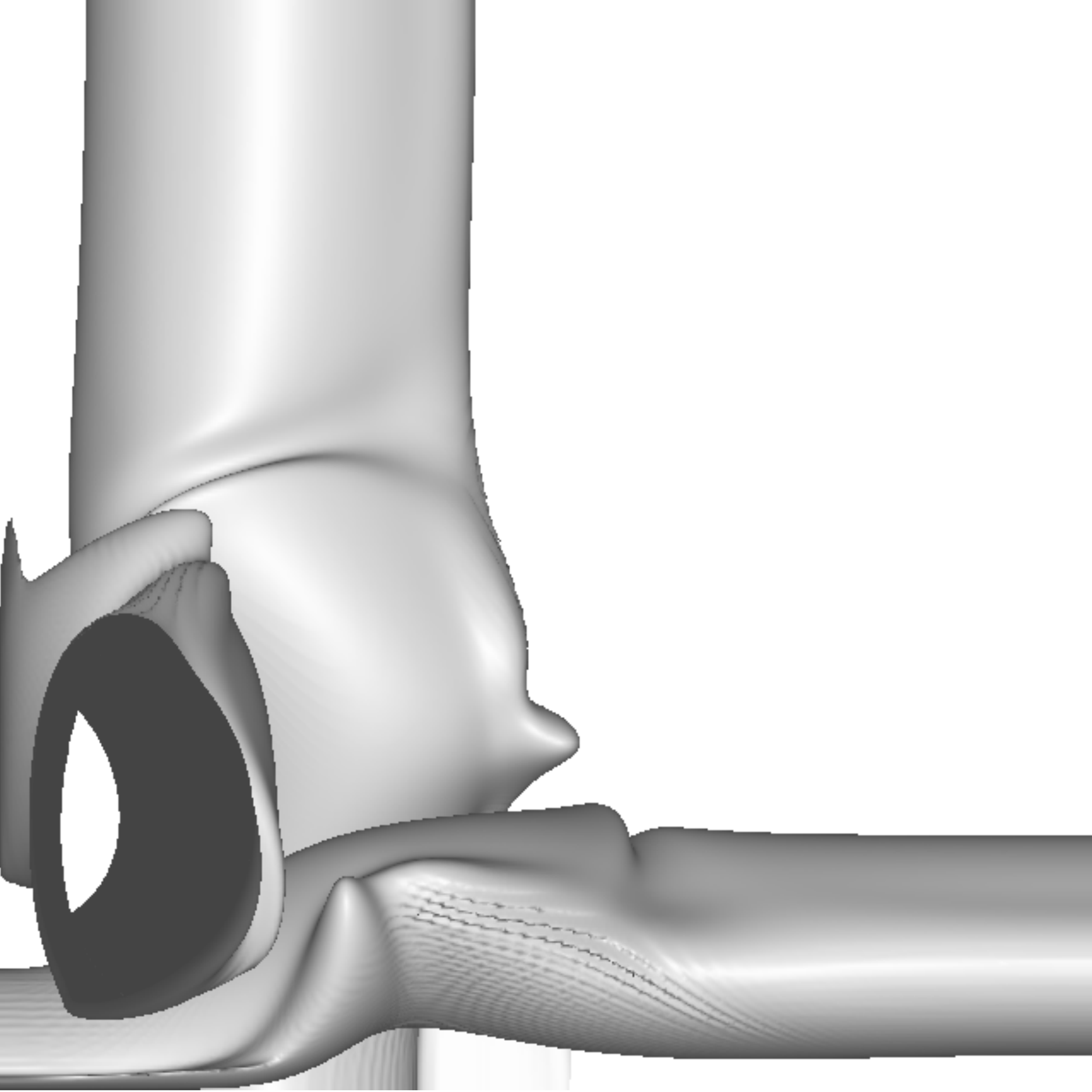} \\
  \vspace*{2mm}
  \includegraphics[height=0.179\textheight]{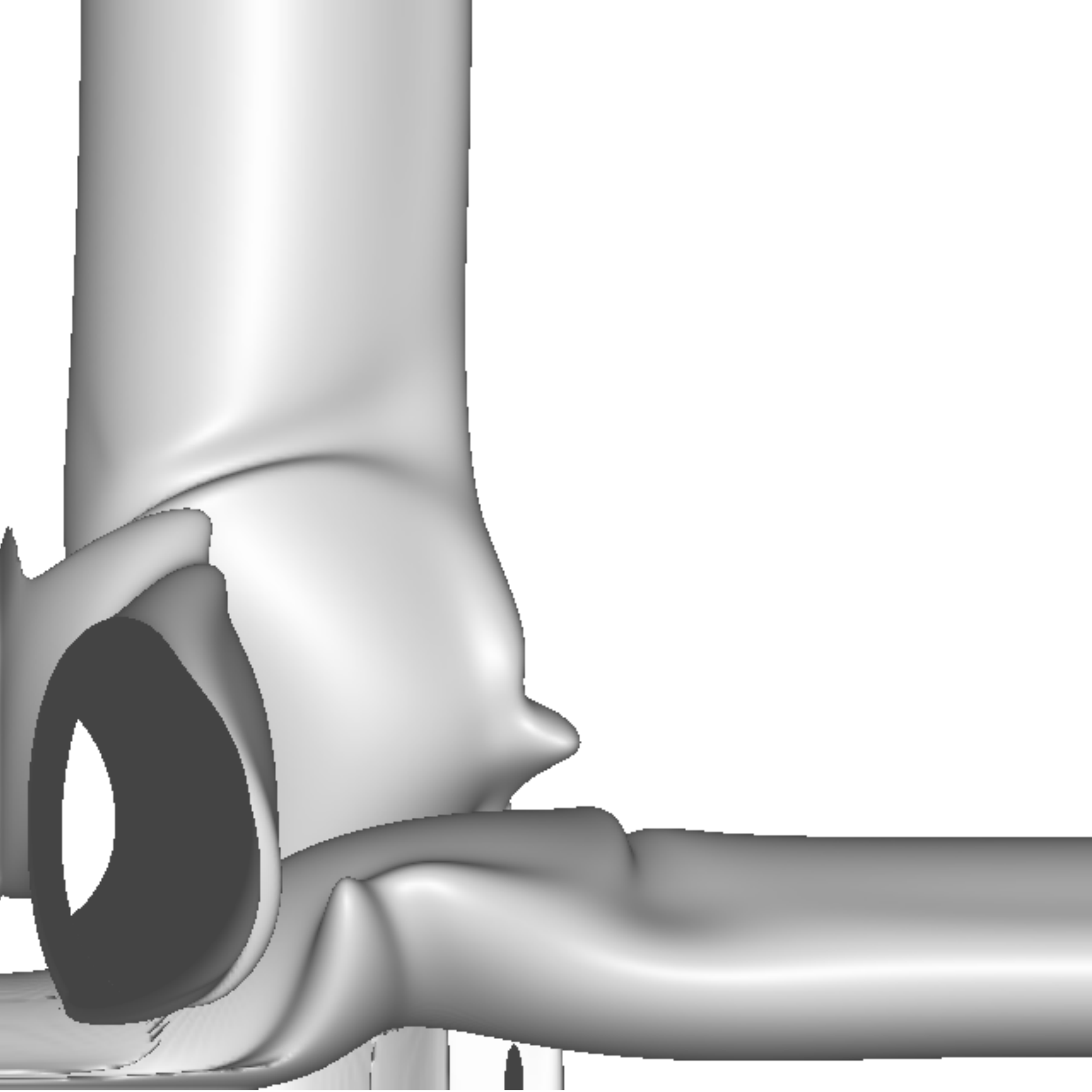} \\
  \vspace*{2mm}
  \includegraphics[height=0.179\textheight]{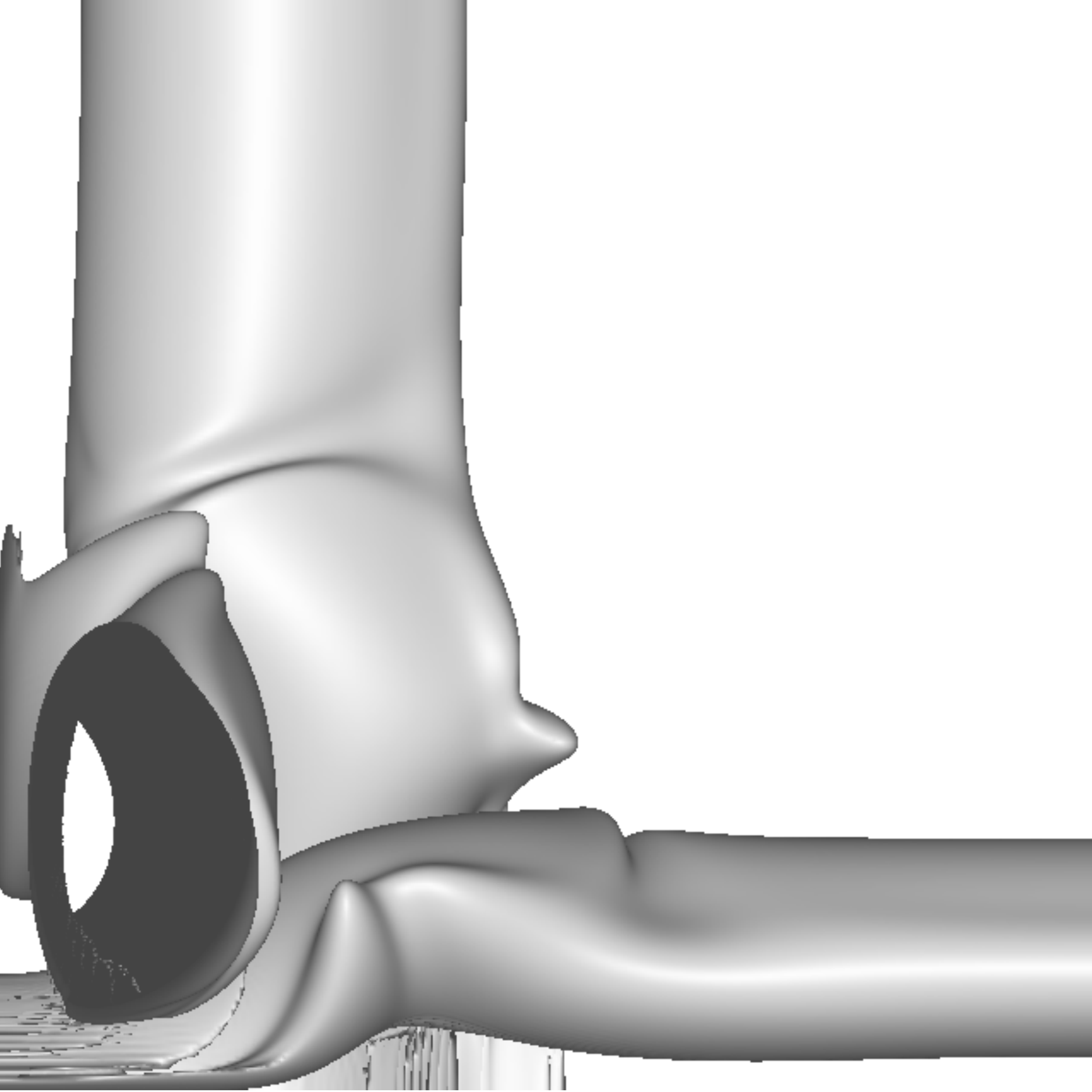}
  \caption{\label{spectral} Isosurface plots of vorticity. From top to bottom:
    spherical truncation, 2/3 rule, exponential filtering,
    Harlow-Welsh, vorticity formulation, $512^3$ mesh points}
  \vspace*{-8mm}
\end{figure}

\begin{figure}[!t]
  \includegraphics[width=1\columnwidth]{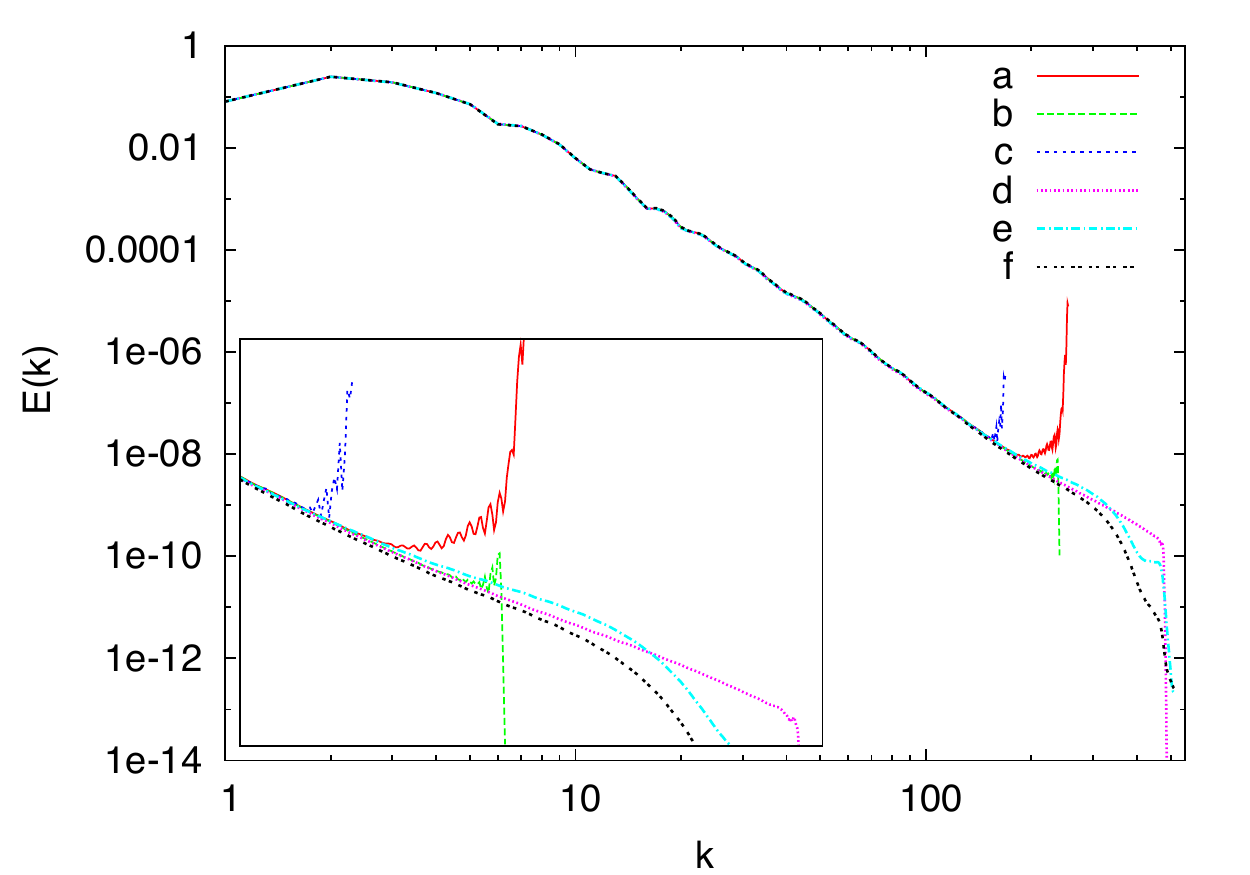}
  \caption{\label{energyspectrum} Energy spectra at time $t = 0.5$ for
    spectral and finite difference methods: a) spherical model truncation ($512^3$),
  b) high-order exponential cut-off ($512^3$), c) $2/3$ rule ($512^3$),
  d) high-order exponential cut-off ($1024^3$), e) vorticity formulation ($1024^3$),
  f) staggered grid formulation ($1024^3$)}
\end{figure}

We first compare the growth of the maximum vorticity according to the
Beale-Kato-Majda result \cite{beale-kato-etal:1984, ponce:1985} for
all six numerical methods described above. The initial condition was
chosen similar to Kida-Pelz 12 vortices \cite{kida:1985,
  boratav-pelz:1994, pelz:2001} with a Gaussian shape for the
vorticity distribution.
Resolution of all the spectral simulations were $512^3$ mesh points
(corresponding to the full domain) and in addition the Hou-Li
exponential filtering was repeated with $1024^3$ mesh points. The
finite difference/volume simulations were performed with $512^3$ and
$1024^3$ mesh points.
The growth of $\max |{\bm\omega}|$ is shown in Fig. \ref{growth_omega}.
All simulation agree very well up to the time when the flow is
underresolved. This is about $t = 0.4$ for the simulations using
$512^3$ mesh points and $t = 0.47$ for the $1024^3$ runs.  There is no
particular criterion which simulation performs better once the
simulation is underresolved. The very simple message from this
comparison is: you just have to resolve the flow and this is more
important than the order of the scheme.

\begin{figure}[!h]
  \includegraphics[width=0.9\columnwidth]{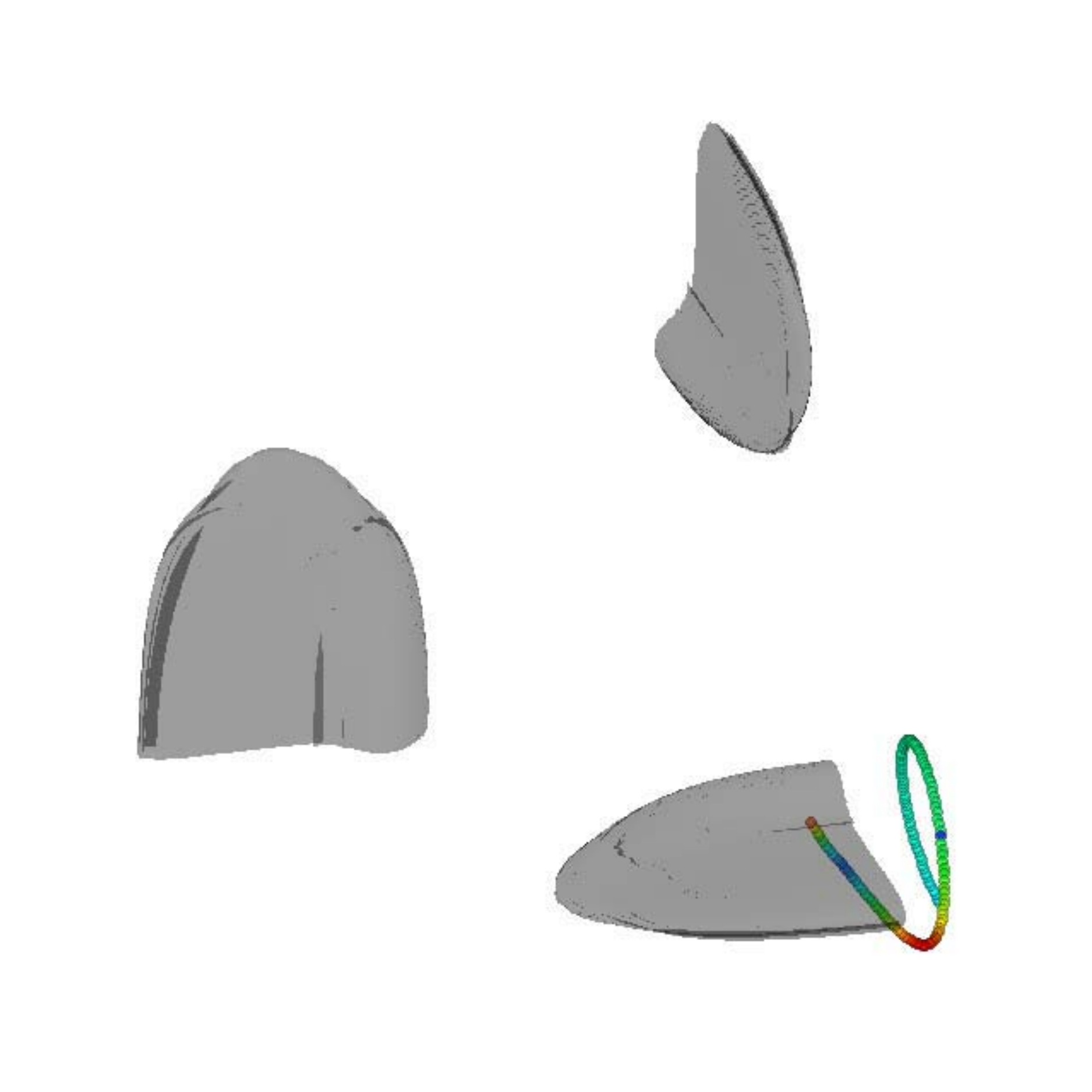}
  \caption{\label{lagrange_particle} Isosurface plot of $\max |{\bm \omega}|$
    at $70\%$ of maximum vorticity. Shown is also the trajectory of a
    particle moving to the position of maximum vorticity.}
\end{figure}

In order to display the differences and similarities of the various
numerical methods, we used a ``low resolution'' simulation with
$512^3$ mesh points at a late time $t=0.5$ where the flow is already
underresolved. Therefore, we looked at very low levels ($5\%$ of the
maximum vorticity) as suggested and done by Kerr \cite{kerr:2006} and Hou and Li
\cite{hou-li:2006}. Due to the high symmetry of the flow, only $1/8$ of
the total configuration is shown. (To get a better impression
for the geometry of the vortices, see Fig. \ref{lagrange_particle},
which shows an isosurface of $70\%$ of the peak vorticity.)

The spherical truncation produces highly visible artifacts due to
heavy oscillations which grow to substantial values. This is mostly
suppressed in the simulation using the classical $2/3$ rule and nearly
vanishes for the high order exponential smoothing. Thus our comparison
confirms the analysis of Hou and Li \cite{hou-li:2006}. Remarkable is
the strong similarity of the real-space methods to the spectral
simulation with high order exponential smoothing. 
This is especially visible in Fig. \ref{energyspectrum}, which shows
the energy spectrum for spectral and finite difference/volume methods
at time $t=0.5$. In the spectral schemes, the spherical truncation and
the 2/3 rule show a strong Gibbs phenomena which is absent in the
exponential filtering and the finite difference/volume schemes. The
Harlow-Welsh method is slightly more dissipative than the vorticity
formulation. From the comparison with the spectral schemes using
exponential filtering and $1024^3$ mesh points it is safe to say
that the finite difference schemes with an approximately 1.3 times
larger resolution in each spatial direction perform equally well as
the spectral code with exponential filtering.
Thus, our conclusions of this comparison is that the differences in
the simulation results caused by the choice of the dealiasing method
are larger than the difference to and between the real-space methods.
Our finding thus confirms the viewpoint of Orlandi and Carnevale
\cite{orlandi-carnevale:2007} and justifies the use of finite
difference/volume methods as integration scheme in an adaptive mesh
refinement treatment.

\begin{figure}[b!]
  \includegraphics[width=1\columnwidth]{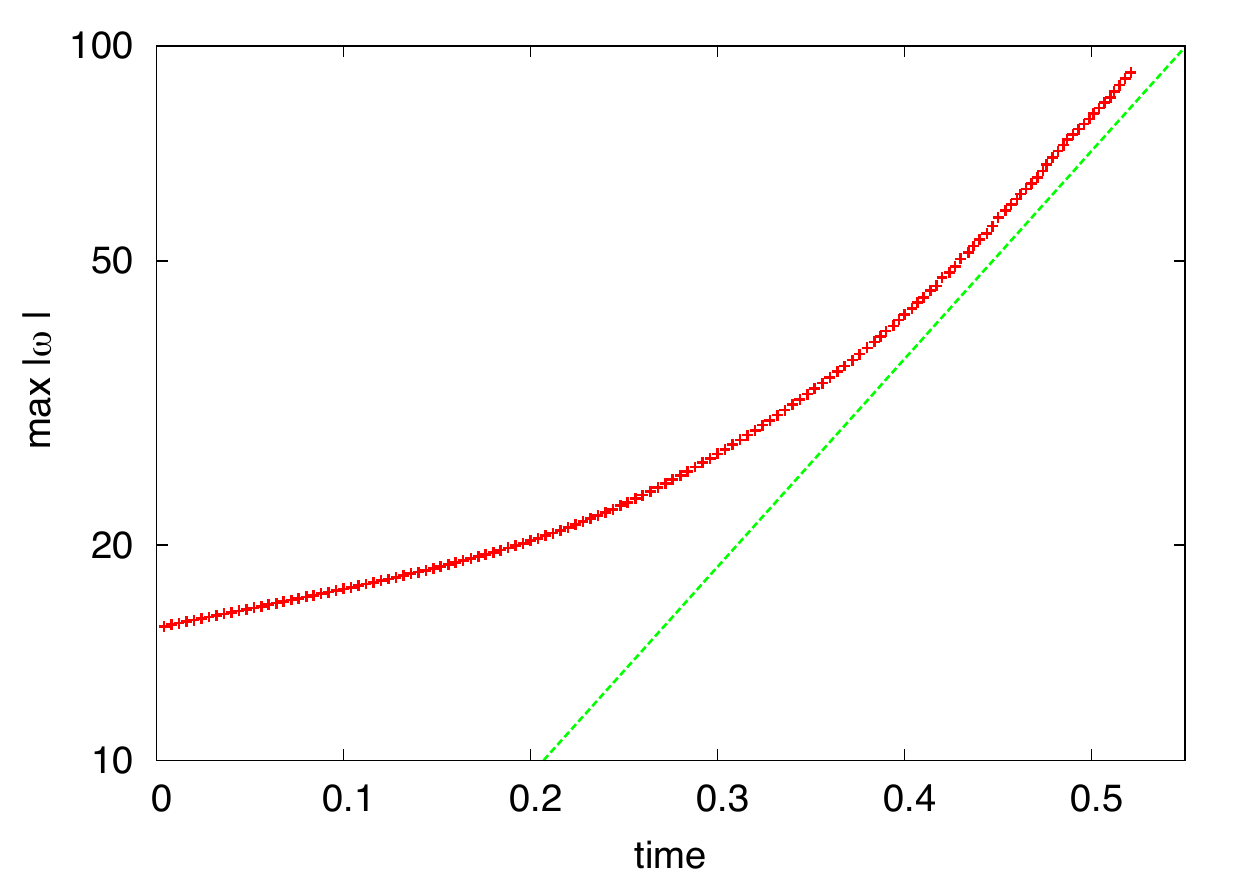}
  \caption{\label{lagrange_growth} Growth of vorticity along the Lagrangian 
    trajectory (red) which ends near the point of maximum vorticity
    and a fitted exponential (green).}
\end{figure}

\subsection{Lagrangian trajectories}

As pointed out in \cite{deng-hou-etal:2005, gibbon:2007}, the
Lagrangian treatment of vorticity amplification is closely related to
the local geometric properties -- like curvature and torsion -- of
vortex lines. In Fig.  \ref{lagrange_particle} the trajectory of a
Lagrangian tracer particle is shown. To obtain this trajectory, we
first identified the spatial position of the maximum vorticity at a
late time of the simulation and then traced back the actual
trajectory. Fig. \ref{lagrange_growth} shows the temporal evolution of
vorticity following this trajectory.  A tendency to an exponential
growth of vorticity along the trajectory is obvious.

\section{\label{amr}Adaptive mesh refinement simulations}

\subsection{The framework \textit{racoon}}

For the adaptive mesh refine calculations, we use our framework
\textit{racoon} \cite{dreher-grauer:2004} which is designed for
massive parallel computations and scales for hyperbolic systems
linearly up to 16384 processors on BlueGene BG/L. However, for the
incompressible Euler equations, the pressure resp.~ vector potential
are solved using an adaptive multigrid method 
\cite{brandt:1977, barad-colella:2005}
which presently
scales only up to 64 processors. Therefore, the present simulations
are limited to an effective resolution of $4096^3$ mesh points.
Parallelization and load-balancing is performed using a space-filling
Hilbert curve 
\cite{dreher-grauer:2004}.

\begin{figure}[!h]
  \includegraphics[width=0.9\columnwidth]{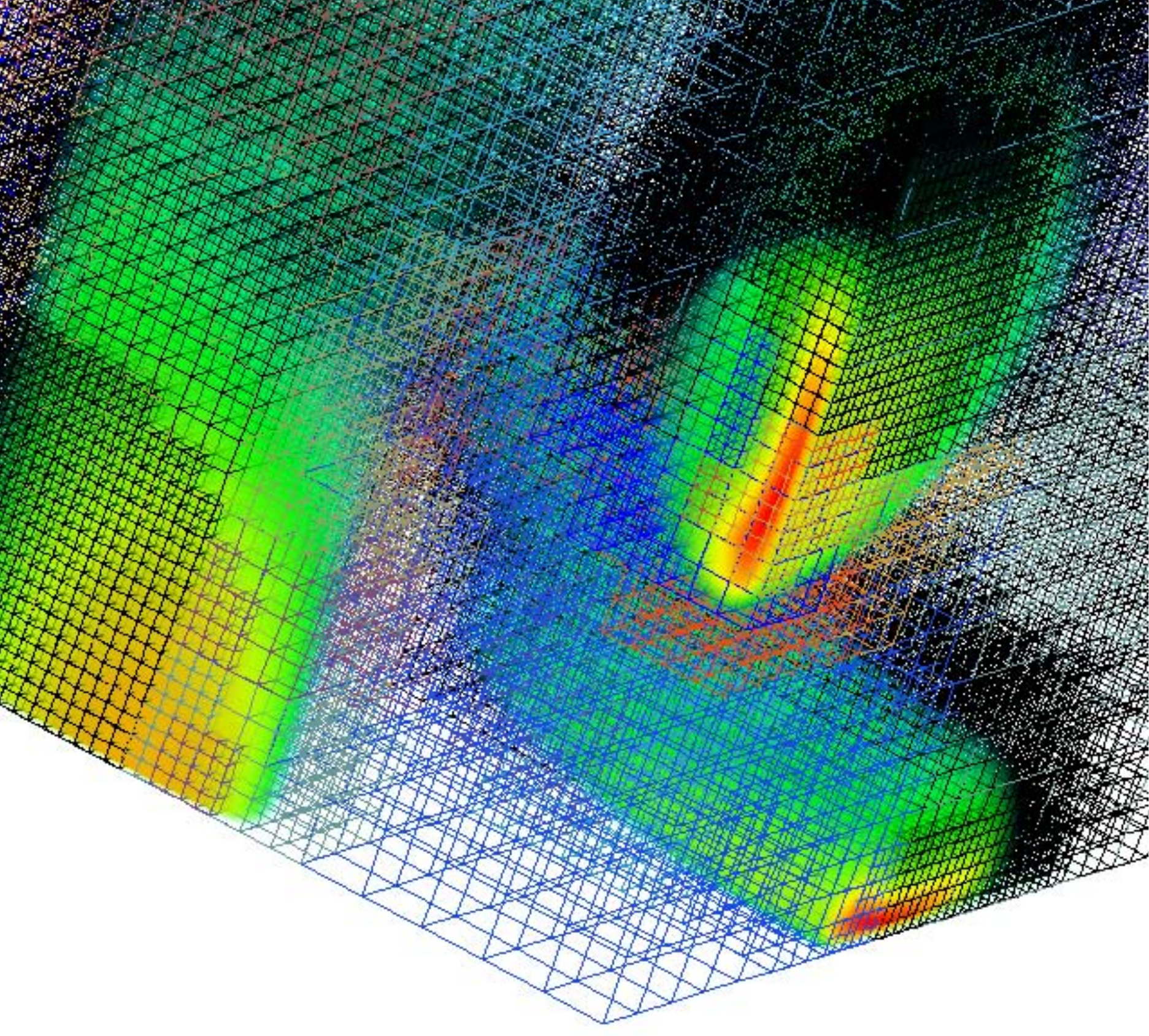}
  \caption{\label{amrtubes}Volume rendering of vorticity at time $t=0.5$.}
  \vspace*{-0.8cm}
\end{figure}

Using the framework racoon and the vorticity formulation, we solve the
incompressible Euler equations with an effective resolution of
$4096^3$ mesh points. Fig. \ref{amrtubes} shows a volume rendering of
vorticity at the latest time $t=0.5$ including the adaptive meshes.
Memory consumption is quite moderate using less than 80 GBytes.

\subsection{Analyzing the growth of vorticity}

\begin{figure}[!t]
  \includegraphics[width=1\columnwidth]{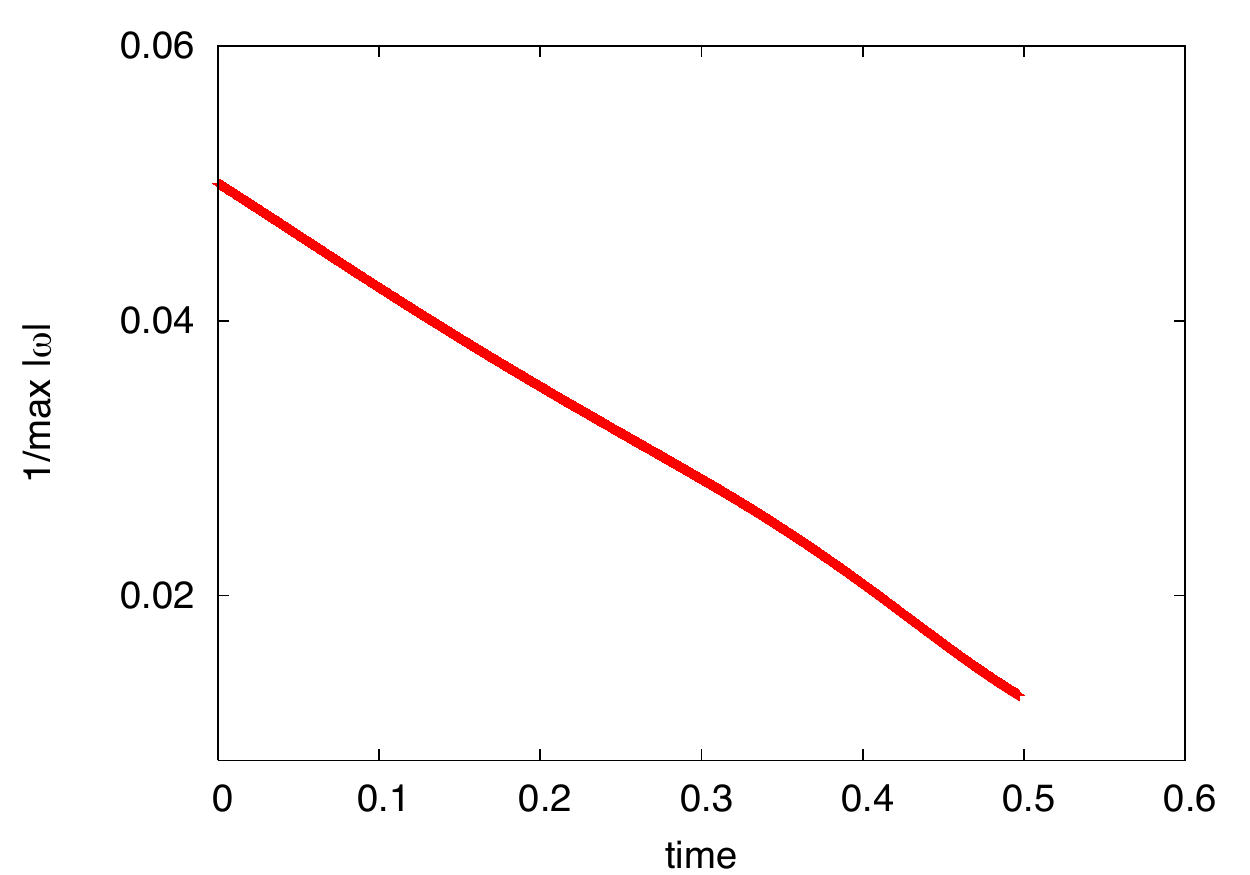}
  \caption{\label{1dom}Temporal evolution of $1/\max |{\bm \omega}|$.}
\end{figure}

Looking at Fig. \ref{1dom} which shows the time evolution of $1/\max
|{\bm \omega}|$ it is tempting to identify a finite time singularity.
However, a more appropriate presentation is obtained plotting $\max
|{\bm \omega}| \times (t_0 - t)$ where $t_0$ is the expected
singularity time.
This quantity should converge to a horizontal line in this plot if a
singularity occurs in finite time. The time $t_0=0.638$ is chosen in a
way that this scaling is observed in the late phase of the simulation
while the numerics is still resolved.
\begin{figure}[!b]
  \includegraphics[width=1\columnwidth]{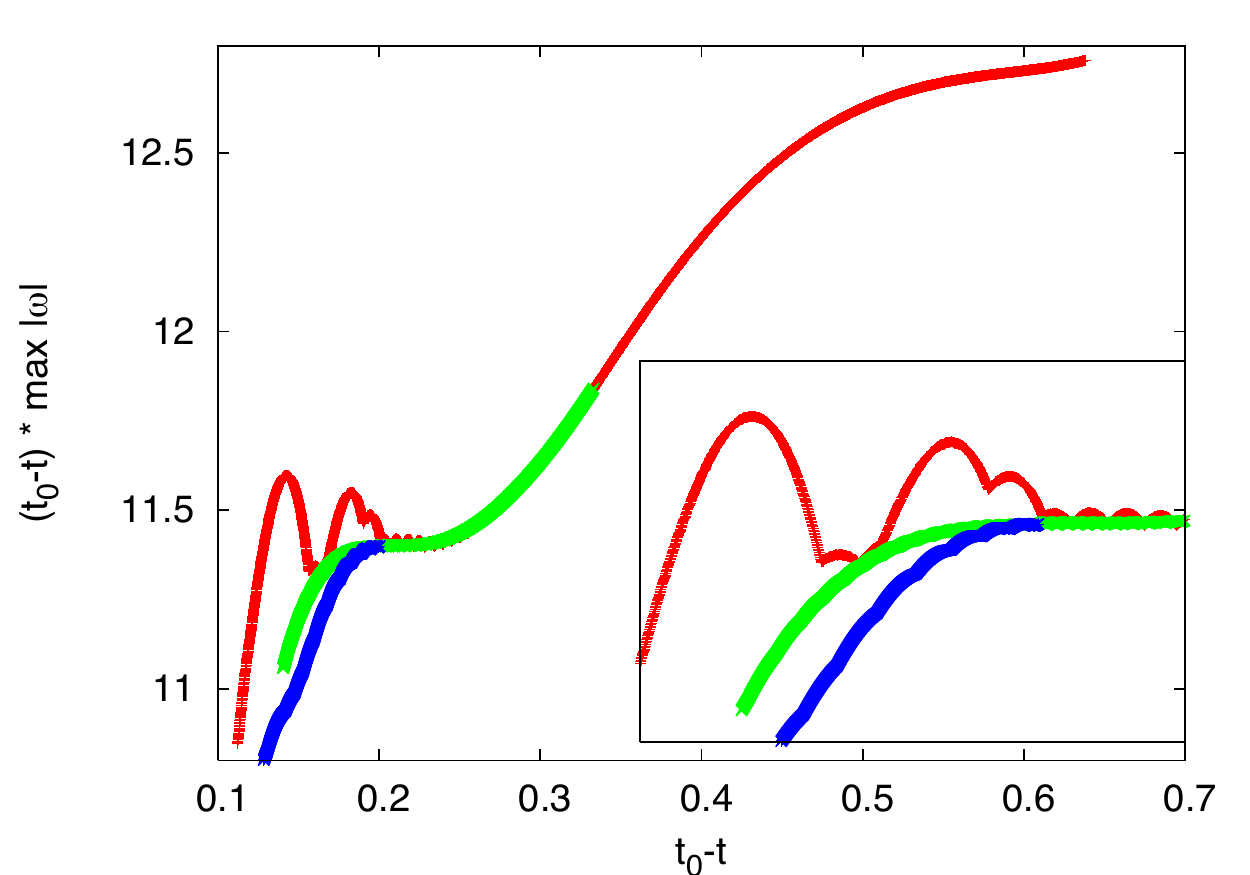}
  \caption{\label{scaling} Scaling of the growth of vorticity.  Red:
    $1024^3$ mesh points, Blue: $2048^3$ mesh points, Green: $4096^3$
    mesh points. The inlet shows the late phase of the simulation
    and highlights the importance of numerical resolution.}
\end{figure}
%
%
This is shown in Figs. \ref{scaling} and the zoom in the inlet of this figure.
Especially the zoom of the late phase of the simulation demonstrates,
how sensitive the growth of vorticity depends on the numerical
resolution and that conclusions drawn from underresolved simulations
must be handled with care.

\section{Conclusions and Outlook}


We demonstrated the extreme sensitivity of the growth of vorticity on
the numerical resolution. In order to gain further insight into the
mechanism of vorticity amplification, future simulations should
include the following analysis and diagnostics:
i) If a finite time singularity is expected, then the blow-up time of
vorticity must occur at the same time when the spatial position of
maximum vorticity and maximum strain come together.
ii) The Lagrangian viewpoint should be analyzed according to Deng, Hou
and Xu \cite{deng-hou-etal:2005} and Gibbon \cite{gibbon:2007}.
iii) Simulations should use initial conditions including the Kida-Pelz
flow \cite{kida:1985} and Bob Kerr's orthogonal tubes
\cite{kerr:1993}. However, the shape of the initial vortex tube should
be chosen in such a way that vortex shedding will not pollute the
vorticity growth. For orthogonal vortex tubes this was achieved by
Orlandi and Carnevale \cite{orlandi-carnevale:2007} starting with Lamb
dipoles.

\begin{acknowledgments}
  R.G. likes to thank Gregory Eyink, John Gibbon, Thomas Hou, Robert
  Kerr and Miguel D. Bustamante for fruitful discussion and Uriel
  Frisch and his coworkers for organizing this conference.
  Access to the JUMP multiprocessor computer at the FZ J\"ulich was
  made available through project HB022. Part of the computations were
  performed on an Linux-Opteron cluster supported by HBFG-108-291.
  
\end{acknowledgments}

\bibliography{EE250}

\end{document}